\documentclass[11pt, leqno, amsfonts]{article}
\usepackage{amsmath,latexsym}

\usepackage{amsmath}

\usepackage[dvips]{graphicx}

\makeatletter
\newcommand{\@makemycaption}[2]{%
\vspace{10pt}%
{\textbf{#1}:#2\par}%
}
\renewcommand{\figure}{%
\let\@makecaption\@makemycaption\@float{figure}}
\makeatother

\date{}

  {\end{list}%
  }%
\newcommand{\bx}{{\bf x}}
\newcommand{\by}{{\bf y}}
\newcommand{\bw}{{\bf w}}
\newcommand{\bv}{{\bf v}}
\newcommand{\bs}{{\bf s}}

\newcommand{\bz}{{\bf z}}
\newcommand{\bu}{{\bf u}}

\newcommand{\wht}{\widehat}
\newcommand{\var}{{\rm Var}}

\newcommand{\cV}{{\cal V}}
\newcommand{\cX}{{\cal X}}

\newcommand{\cov}{{\rm Cov}}

\begin{document}
\begin{center}
{\LARGE Importance Sampling of Word Patterns  }

{\LARGE in DNA and Protein Sequences }
\medskip

HOCK PENG CHAN$^\dag$

{\it Department of Statistics and Applied Probability, National University of Singapore, Singapore 119260, Republic of Singapore \\}
{\it tel:(65)65166750 fax:(65)68723919 email:stachp@nus.edu.sg}

\medskip

NANCY RUONAN ZHANG$^{*\dag}$

{\it Department of Statistics, Stanford University, Stanford,
CA 94305-4065, USA\\}
{\it tel:(650)7232620 fax:(650)7258977 email:nzhang@stanford.edu}

\medskip

LOUIS H.Y. CHEN

{\it Institute for Mathematical Sciences, National University of Singapore,
Singapore 118402, Republic of Singapore \\}
{\it tel:(65)65161897 fax:(65)68738292 email:matchyl@nus.edu.sg}

\medskip
\today

\medskip

{\bf ABSTRACT}
\end{center}

\noindent {\bf Monte Carlo methods can provide accurate p-value estimates of word
counting test statistics and are easy to implement. They are
especially attractive when an asymptotic
theory is absent or when either the search sequence or the word pattern
is too short for the application of asymptotic formulae. Naive direct Monte
Carlo is undesirable for the estimation of
small probabilities because the associated rare events of interest are seldom generated.
We propose instead efficient importance sampling algorithms that use controlled insertion
of the desired word patterns on randomly generated sequences.
The implementation is illustrated on word patterns of biological
interest: Palindromes and inverted repeats, patterns
arising from position specific weight matrices and co-occurrences
of pairs of motifs. }

\smallskip


\smallskip
\noindent $^\dag$ joint first authors.

\noindent $^*$ corresponding author.

\newpage
\def\theequation{1.\arabic{equation}}
\setcounter{equation}{0}
\begin{center}
{\bf 1. INTRODUCTION}
\end{center}

\smallskip \noindent
Searching for matches to a word pattern, also called a motif,
is an important task in computational biology. The word pattern
represents a functional site, such as a transcription
factor binding site (TFBS) in a promoter region of
a DNA sequence or a ligand docking site in a protein sequence.
Statistical significance of over-representation of these word patterns
provides valuable clues to biologists and as a result, there have been a lot of work done
on the use of asymptotic limiting distributions to approximate these
p-values, see Prum et al. (1995), Reinert et al. (2000), R\'{e}gnier (2000), Robin et al. (2002),
Huang et al. (2004), Leung et al. (2005),
Mitrophanov and Borodovsky (2006), Pape et al. (2008) and references therein.
However, the approximations may not be accurate for short words or for words consisting of
repeats and most theoretical approximations work only in specific settings.
String-based recursive methods can provide exact p-values, see for example Gusfield (1997), but they can be computationally expensive when the number of words in the word pattern
is large.

Direct Monte Carlo algorithms for estimating p-values of word patterns are easy to
implement but are inefficient for the estimation of very small
p-values because in such cases, almost all the simulated sequences do not contain the
required number of word patterns. We propose in this paper importance sampling
algorithms that insert the desired word patterns either randomly or controlled by a
hidden Markov model, on the simulated sequences. The algorithms are described
in Section 2 and are illustrated on several word patterns of biological interest:
Palindromes and inverted repeats in Section 3, high-scoring words
with respect to position specific weight matrices in Section 4 and
co-occurrences of motifs in Section 5. Numerical results show that variance reduction of several
orders of magnitude are achieved when applying the proposed importance sampling algorithms on
small p-values. The technical details are consolidated in the appendices and include a proof
of the asymptotic optimality of the importance sampling algorithms, in Appendix D.

\newpage
\def\theequation{2.\arabic{equation}}
\setcounter{equation}{0}
{\bf 2. IMPORTANCE SAMPLING OF WORD PATTERNS}

\smallskip
\noindent {\it 2.1 Word counting}

\smallskip Let $|B|$ denote the
number of elements in a set $B$. By selecting
randomly from a finite set $B$, we shall mean that each $b \in B$ has probability
$|B|^{-1}$ of being selected. For any two sequences
$\bv = v_1 \cdots v_m$ and $\bu = u_1 \cdots u_r$, the notation
$\bv \bu$ shall denote the concatenated sequence $v_1 \cdots v_m u_1 \cdots u_r$.
We also denote the length of $\bv$ by $\ell(\bv)(=m)$. Although we assume implicitly
an alphabet $\cX = \{ a,c,g,t \}$, representing the four nucleotide bases of DNA sequences,
the algorithms can be applied on any countable alphabet,
for example the alphabet of 20 amino acids in protein sequences.

We will represent the word pattern of interest by a set of words $\cV$ and assume that $|\cV|
< \infty$. Let $\bs = s_1 \cdots s_n$ denote a
sequence of DNA bases under investigation
and let $N_m$ be the maximum number of non-overlapping
words from $\cV$ in $\bs_m = s_1
\cdots s_m$. We say that there exists a word in $\cV$ at the end of $\bs_m$
if $s_{m-j+1} \cdots s_m \in \cV$ for some $j > 0$. Moreover, the smallest
such $j$ is the length of the shortest word at the end of $\bs_m$.
We have the recursive relations, for $m \geq 1$,
\begin{equation} \label{Nm}
N_m = \begin{cases} N_{m-1} & \mbox{ if there is no word in } \cV \mbox{ at
the end of } \bs_m, \cr
N_{m-j}+1 & \mbox{ if the shortest word in } \cV \mbox{ at the end of }
\bs_m \mbox{ is of length } j,
\end{cases}
\end{equation}
with the initialization $N_0=0$. We denote $N_n$ simply by $N$. It is also possible to
modify (\ref{Nm})
to handle the counting of possibly overlapping words.

\bigskip
\noindent {\it 2.2 Monte Carlo evaluation of statistical significance}

\smallskip We begin by describing direct Monte Carlo. To evaluate the
signifiance of observing $c$ word patterns in an observed sequence $\bs$,
we generate independent copies of the sequence from a Markov chain
with transition probabilities estimated either from $\bs$ or from a local neighborhood of $\bs$.
The proportion of times $\{ N \geq c \}$ occurs among the independent copies of $\bs$
is then the direct Monte
Carlo estimate of the p-value $p_c:=P \{ N \geq c \}$.

It is quite common for many sequences to be analyzed simultaneously. Hence
to correct for the effect of multiple comparisons, a very small p-value is required
for any one sequence before
statistical significance can be concluded. Direct Monte Carlo is well-known to be very
inefficient for estimating small probabilities in general and many
importance sampling schemes have been proposed to overcome this drawback, for example
in sequential
analysis (Siegmund, 1976), communication systems (Cottrell, Fort and Malgouyres, 1983),
bootstrapping (Johns, 1988 and Do and Hall, 1992), signal detection
(Lai and Shan, 1999), moderate deviations (Fuh and Hu, 2004)
and scan statistics (Chan and Zhang, 2007). In this paper,
we provide change of measures that are effective
for the importance sampling of word patterns.

For ease of exposition, assume that the background sequence of bases follows a first-order
Markov chain with positive transition probabilities
\begin{equation}
\label{Psi}
\sigma(xy) := P \{ s_{i+1}=y|s_i=x \}, \quad x,y \in \cX.
\end{equation}
Let $\pi$ be
the stationary distribution and let
$\sigma(v_1 \cdots v_i) = \prod_{j=1}^{i-1}
\sigma(v_j v_{j+1})$. Before executing the importance sampling algorithms, we first create
a word bank of the desired word
pattern, with each word in the word bank taking the value $\bv \in \cV$ with probability
$q(\bv) > 0$. The procedure for the selection of $q$ and construction of the word
banks will be elaborated in Sections 3--5. For completeness, we define $q(\bv)=0$ when
$\bv \not\in \cV$. Let $\beta(\bv) = q(\bv)/
\sigma(\bv)$. For ease of computation, we shall generate a dummy variable $s_0$
before generating $\bs$ and denote $s_0 \cdots s_n$ by $\bs_0$. The first
importance sampling algorithm, for the estimation of $p_1$ only, is as follows.

\smallskip
{\bf ALGORITHM A} (for $c=1$).

\begin{enumerate}
\item Select a word $\bv$ randomly from the word bank. Hence the word takes the value $\bv \in
\cV$ with probability $q(\bv)$.

\item Select $i_0$ randomly from $\{ 1,\ldots,n-\ell(\bv)+1 \}$.

\item Generate $s_0$ from the stationary distribution
and $s_1$, $\ldots$, $s_{i_0-1}$ sequentially from the underlying Markov chain.
Let $s_{i_0} \cdots s_{i_0+\ell(\bv)-1}
=\bv$ and generate $s_{i_0+\ell(\bv)}$, $\ldots$, $s_n$ sequentially from the underlying
Markov chain.
\end{enumerate}

\smallskip
Let $\ell_{\min} = \min_{\bv \in \cV} \ell(\bv)$ and $\ell_{\max} =
\max_{\bv \in \cV} \ell(\bv)$. Recall that $\beta(\bv)=0$ for $\bv \not\in \cV$. Then
\begin{equation}
\label{Lsk}
L(\bs_0) := \sum_{\ell=\ell_{\min}}^{\ell_{\max}} (n-\ell+1)^{-1}
\sum_{i=1}^{n-\ell+1} \beta(s_i \cdots s_{i+\ell-1})/\sigma(s_{i-1} s_i)
\end{equation}
is the likelihood ratio of generating $\bs_0$ from Algorithm A and from the underlying Markov
chain (with no insertion of word patterns). If Algorithm A is run independently
$K$ times, with the $k$th copy of $\bs_0$ generated denoted by
$\bs_0^{(k)}$, then
\begin{equation} \label{pI}
\widehat p_{\rm I} := K^{-1} \sum_{k=1}^K L^{-1}(\bs_0^{(k)}) {\bf 1}_{\{ N^{(k)}
\geq c \}}
\end{equation}
is unbiased for $p_c$. The term ${\bf 1}_{\{ N^{(k)}
\geq c \}}$ is superfluous when using Algorithm A
since at least one word pattern from $\cV$ is generated in
every copy of $\bs_0$.

We restrict Algorithm A to $c=1$ because the random insertion of more than one word patterns
into the simulated sequence can result in a hard to compute likelihood
ratio. To handle more general $c$, we use a hidden Markov model device in Algorithm B
below, with hidden states $X_i$ taking
either value 0 (do not insert word pattern) or 1 (insert word pattern), so that
the likelihood ratio can be computed recursively. Let
\begin{equation} \label{rh}
\rho_i = P \{ X_i = 1 | s_0 \cdots s_i \}
\end{equation}
be the word insertion probability at position $i+1$ along the DNA sequence.
For example, the user can simply select $\rho_i=c/n$ for all $i$
so that approximately $c$ word patterns are inserted in each generated sequence $\bs_0$.
Each copy of $\bs_0$ is generated in the following manner.

\smallskip
{\bf ALGORITHM B} (for $c \geq 1$).

\begin{enumerate}
\item  Let $i=0$, generate $s_0$ from the stationary distribution and $X_0$
satisfying (\ref{rh}).

\item \begin{enumerate}
\item If $X_i=1$, select a word $\bv$ randomly from the word bank. If $\ell(\bv)
\leq n-i$, that is, if the word $\bv$ can fit into the remaining sequence, let
$s_{i+1} \cdots s_{i+\ell(\bv)}=
\bv$, generate $X_{i+\ell(\bv)}$ according to (\ref{rh}), increment $i$ by $\ell(\bv)$
and go to step 3.

\item If the word selected in 2(a) cannot fit into the remaining sequence or if $X_i=0$,
generate $s_{i+1}$ from the underlying Markov chain and $X_{i+1}$ satisfying
(\ref{rh}). Increment $i$ by 1 and go to step 3.
\end{enumerate}

\item If $i < n$, repeat step 2. Otherwise, end the recursion.
\end{enumerate}

\smallskip
Let $L_i=L_i(s_0 \cdots s_i)$ be the likelihood ratio of generating
$s_0 \cdots s_i$ from Algorithm B and from the underlying Markov chain.
Let $\gamma_j = \sum_{\bv \in \cV: \ell(\bv) \leq j} q(\bv)$
be the probability that a randomly chosen word from the word bank
has length not exceeding $j$. Then
\begin{equation}
\label{Lis}
L_i = (1-\rho_{i-1} \gamma_{n-i+1}) L_{i-1} + \sum_{\ell=\ell_{\min}}^{\ell_{\max}}
\rho_{i-\ell} L_{i-\ell} \beta(s_{i-\ell+1} \cdots s_i) /\sigma(s_{i-\ell} s_{i-\ell+1})
\mbox{ if } i \geq 1,
\end{equation}
with $L_i=0$ for $i \leq 0$.

The estimator (\ref{pI}), with $L=L_n$, is unbiased if and only if all configurations of $\bs_0$
satisfying $N \geq c$ can be generated via Algorithm B. To ensure this, it suffices for us to
impose the constraint
\begin{equation} \label{rho}
\rho_i < 1 \mbox{ for all } i < n-\ell_{\min}(c-N_i),
\end{equation}
so that we do not force the insertion of too many word patterns.

\newpage
\def\theequation{3.\arabic{equation}}
\setcounter{equation}{0}
\begin{center}
{\bf 3. PALINDROMIC PATTERNS AND INVERTED REPEATS}
\end{center}

\smallskip Masse et al. (1992) reported clusters of palindromic
patterns near origin of replications of viruses. There have been much work done to estimate
their significance, for example using Poisson and compound Poisson
approximations, see Leung et al. (1994,
2005). The four nucleotides can be divided into two complementary base pairs with $a$
and $t$ forming a pair
and $c$ and $g$ forming the second pair. We denote this relation by writing
$a^c=t$, $t^c=a$, $c^c=g$ and
$g^c=c$. For a word $\bu_m=u_1 \cdots u_m$, we define its complement $\bu_m^c=u_m^c \cdots
u_1^c$. A palindromic pattern of length $\ell=2m$ is a DNA sequence that can be expressed
in the form $\bu_m \bu_m^c$. For example, $\bv=acgcgt$ is a palindromic pattern. Note that the
complement of $\bv$, that is the word obtained by replacing each letter of $\bv$
by its complement, is $tgcgca$,
which is just $\bv$ read backwards. This interesting property explains
the terminology ``palindromic pattern''.

Inverted repeats can be derived from palindromic patterns by inserting a DNA
sequence of length $d$ in the exact
middle of the pattern. The class of word patterns for inverted repeats can be expressed in the
form
\begin{equation}
\label{invert}
\cV = \{ \bu_m \bz \bu_m^c: d_1 \leq \ell(\bz) \leq d_2 \},
\end{equation}
with $0 \leq d_1 \leq d_2$. When $d_1=d_2=0$, then (\ref{invert}) is
the class of all palindromic patterns of length $2m$.

The construction of word banks for palindromic patterns is straightforward. It all boils
down to generating $\bu_m$ in some suitable manner. We advocate generating $\bu_m$
with probability proportional to $\pi(u_1) \sigma(\bu_m) \sigma(\bu_m^c)$ or $\pi(u_1)
\sigma(\bu_m \bu_m^c)$ and show how this can be done in Appendix A.

Having a word bank for palindromic patterns allows us to create a word bank for inverted repeats
easily. The procedure is as follows.

\begin{enumerate}
\item Select $\bu_m \bu_m^c$ randomly from a word bank of palindromic patterns and $d$
randomly from $\{ d_1, \ldots, d_2 \}$.

\item Let $z_0 =u_m$ and
generate $z_1, \ldots, z_d$ sequentially from the underlying Markov chain.

\item Store the word $\bu_m \bz_d \bu_m^c$ into the word bank for inverted repeats.
\end{enumerate}

This procedure allows $\gamma_j$, see (\ref{Lis}), to be computed easily. In particular,
$\gamma_j= (j-d_1+1)/(d_2-d_1+1)$ for $d_1 \leq j \leq d_2$, $\gamma_j=0$ for
$j<d_1$ and $\gamma_j=1$ for $j>d_2$.

\newpage

\def\theequation{4.\arabic{equation}}
\setcounter{equation}{0}
\begin{center}
{\bf 4. POSITION SPECIFIC WEIGHT MATRIX (PSWM)}
\end{center}

\smallskip PSWMs are commonly used to derive fixed-length word patterns or motifs
that transcription factors bind onto and usually range from four to
twenty bases long.. Databases such as TRANSFAC, JASPAR and SCPD
curate PSWMs for families of transcription factors.
For example, the PSWM for the SWI5 transcription factor in the yeast genome is
\begin{equation} \label{swi5} \begin{matrix} a \cr c \cr g \cr t \end{matrix}
\left( \begin{array}{cccccccccccc} 4 & 0 & 4 & 1 & 1 & 4 & 0 & 0 & 0 & 0 & 0 & 2 \cr
                     1 & 2 & 1 & 1 & 3 & 2 & 0 & 0 & 7 & 0 & 0 & 0 \cr
                     2 & 2 & 0 & 2 & 1 & 0 & 2 & 7 & 0 & 0 & 7 & 5 \cr
                     0 & 3 & 2 & 3 & 2 & 1 & 5 & 0 & 0 & 7 & 0 & 0 \end{array} \right),
\end{equation}
see Zhu and Zhang (1999).
Let $w_i(v)$ denote the entry in a PSWM that corresponds to base $v$ at column
$i$ and let $m$ be the number of columns in the PSWM. For any word $\bv_m$ (of length $m$),
a score
$$S(\bv_m) := \sum_{i=1}^m w_i(v_i)
$$
is computed and words with high scores are of interest. We let $\cV$ be the set of all
$\bv_m$ with score not less than
a pre-specified threshold level $t$. In other words,
\begin{equation} \label{VPSWM} \cV = \{ \bv_m: S(\bv_m) \geq t \}
\end{equation}
is a motif for the PSWM associated with a given transcription factor. The matrix is derived from
the frequencies of the four bases at various positions of known instances of the TFBS,
which are usually confirmed by biological experiments. Huang et al. (2004) gave a good review
of the construction of PSWMs.

In principle, we can construct a word bank for $\cV$ by simply generating words
of length $m$ from the underlying Markov chain and discarding words that do not
belong to the motif. However for $t$ large, such a procedure involves
discarding a large proportion of the generated words. It is more efficient to
generate the words with a bias towards larger scores.
In Appendix B, we show how, for any given $\theta > 0$, a tilted Markov
chain can be constructed to generate words $\bv$ with probability mass function
\begin{equation} \label{tilted}
q_\theta(\bv) = e^{\theta S(\bv)} \pi(v_1) \sigma(\bv)/ \Lambda(\theta),
\end{equation}
where $\Lambda(\theta)$ is a computable normalizing constant. If words
with scores less than $t$ are discarded, then the
probability mass function of non-discarded words is
\begin{equation} \label{tq}
q(\bv) = \xi e^{\theta S(\bv)} \pi(v_1) \sigma(\bv)/\Lambda(\theta)
\mbox{ for } \bv \in \cV,
\end{equation}
where $\xi$ is an unknown normalizing constant that can be estimated
by the reciprocal of the fraction of non-discarded words.
There are two conflicting demands placed on the choice of $\theta$. As $\theta$ increases, the
expected score of words generated under $q_\theta(\bv)$ increases. We would thus
like $\theta$ to be large so that the fraction of discarded words
is small. However at the same time, we would also like
$\theta$ to be small, so that the variation of $\beta(\bv)=q(\bv)/\sigma(\bv)$
over $\bv \in \cV$ is small. Since
\begin{equation} \label{qstar}
E_{q_\theta}[S(\bv)] = \frac{d}{d \theta} [\log \Lambda(\theta)],
\end{equation}
we suggest choosing the root of the equation $\frac{d}{d \theta} [\log
\Lambda(\theta)]=t$. See Appendix B for more details on the the computation of $\Lambda(\theta)$
and the numerical search of the root.

\smallskip \noindent
{\it 4.1 Example 1}

\smallskip We illustrate here the need for alternatives to analytical p-value
approximations by applying Algorithm A on some special word patterns. Let $P_\pi$ denotes
probability with $v_1$ following stationary distribution $\pi$. Huang et al. (2004) suggested
an approximation, which for $c=1$ reduces to
\begin{equation} \label{PN}
P \{ N \geq 1 \} \doteq 1-(1-P_\pi \{ S(\bv_m) \geq t \})^{n-m+1}.
\end{equation}

Consider $s_1,\ldots,s_n$ independent and identically distributed random variables taking
values $a$, $c$, $g$ and $t$ with equal probabilities. Let
\begin{equation} \label{wrep}
W_{\rm rep} = \begin{matrix} a \cr c \cr g \cr t \end{matrix}
\left( \begin{array}{cccccccccccc} 1 & 1 & 1 & 1 & 1 & 1 & 1 & 1 & 1 & 1 & 1 & 1 \cr
                     0 & 0 & 0 & 0 & 0 & 0 & 0 & 0 & 0 & 0 & 0 & 0 \cr
                     0 & 0 & 0 & 0 & 0 & 0 & 0 & 0 & 0 & 0 & 0 & 0 \cr
                     0 & 0 & 0 & 0 & 0 & 0 & 0 & 0 & 0 & 0 & 0 & 0 \end{array} \right),
\end{equation}
\begin{equation} \label{wnorep}
W_{\rm norep} = \begin{matrix} a \cr c \cr g \cr t \end{matrix}
\left( \begin{array}{cccccccccccc} 1 & 0 & 0 & 0 & 0 & 0 & 0 & 1 & 1 & 0 & 0 & 0 \cr
                     0 & 1 & 0 & 0 & 0 & 0 & 1 & 0 & 0 & 1 & 0 & 0 \cr
                     0 & 0 & 1 & 0 & 0 & 1 & 0 & 0 & 0 & 0 & 1 & 0 \cr
                     0 & 0 & 0 & 1 & 1 & 0 & 0 & 0 & 0 & 0 & 0 & 1 \end{array} \right),
\end{equation}
and consider counting of words with score at least $t$ for $t=9,10$ and 11. The approximation
(\ref{PN}) is the same for both (\ref{wrep}) and (\ref{wnorep}) but we know that the
p-value when the PSWM is (\ref{wrep}) should be smaller due to the tendency of the word
patterns to clump together.
Of course, declumping corrections can be applied to this special case but this
is not so straightforward for general PSWMs. Table 1 compares the analytical, direct Monte
Carlo and importance sampling approximations of $P \{ N \geq 1 \}$ for (\ref{wrep}) and
(\ref{wnorep}) with $n=200$. The simulations reveal substantial over-estimation of p-values for
$W_{\rm rep}$ when using (\ref{PN}). Algorithm A is able to maintain its accuracy over the range
of $t$ considered whereas direct Monte Carlo has acceptable accuracy only for $t=9$.

\smallskip \noindent
{\it 4.2 Example 2}

\smallskip We implement Algorithm B here with
\begin{equation} \label{rhoi}
\rho_i = \min \Big\{ 1, \Big( \frac{c-N_i}{n-i-(c-N_i)(m-1)} \Big)^+ \Big\},
\end{equation}
where $x^+ = \max \{ 0,x \}$. We choose $\rho_i$ in this manner to encourage word insertion
when there are few bases left to be generated and the desired number of word patterns has
not yet been observed. The motif consists of
all words of length 12 having score at least 50 with respect to
the PSWM (\ref{swi5}). The transition matrix for generating the DNA sequence is
\begin{equation} \label{tmat}
\begin{matrix} a \cr c \cr g \cr t \cr \end{matrix}
\left( \begin{matrix} .3577 & .1752 & .1853 & .2818 \cr
.3256 & .2056 & .1590 & .3096 \cr
.2992 & .2180 & .2039 & .2789 \cr
.2381 & .1943 & .1905 & .3771 \cr
\end{matrix} \right),
\end{equation}
and the length of the sequence investigated is $n=700$. We see from Table 2
variance reduction of 10--100 times in the simulation
of probabilities of order $10^{-1}$ to $10^{-3}$.
For smaller probabilities, direct Monte Carlo does not provide an estimate
whereas estimates from the importance sampling algorithm retain their accuracy.
Although importance sampling takes about two times the computing time
of direct Monte Carlo for each simulation run, the savings in
computing time to achieve the same level of accuracy are quite substantial.

\newpage

\def\theequation{5.\arabic{equation}}
\setcounter{equation}{0}
\begin{center}
{\bf 5. CO-OCCURRENCES OF MOTIFS}
\end{center}
\smallskip \noindent
For a more detailed sequence analysis of promoter regions, one can search
for cis-regulatory modules (CRM) instead of single motifs. We define CRM to be
a collection of fixed length motifs that are located in a fixed order in proximity to each
other. They are signals for co-operative binding of transcription factors,
and are important in the study of combinatorial regulation of genes. CRMs have
been used successfully to
gain a deeper understanding of gene regulation, cf. Chiang et al.
(2003), Zhou and Wong (2004) and Zhang et al. (2007). We focus here on the simplest type of CRM:
A co-occurring pair of high scoring words separated by a gap sequence of variable length.
Let $S_1(\cdot)$ be the score of a word of length $m$ calculated
with respect to a PSWM $W_1$, and $S_2(\cdot)$ the score of a word of length $r$ calculated
with respect to a PSWM $W_2$. Let $0 \leq d_1 <
d_2 < \infty$ be the prescribed limits of the length of the gap and $t_1$,
$t_2$ threshold levels for $W_1$ and $W_2$ respectively. The family of words for the
co-occurring motifs is
\begin{equation} \label{v31}
\cV = \{ \bv_m \bz \bu_r: S_1(\bv_m) \geq t_1, S_2(\bu_r)
\geq t_2, d_1 \leq \ell(\bz) \leq d_2 \}.
\end{equation}

In Section 4, we showed how word banks for the motifs $\cV_1 := \{ \bv_m: S_1(\bv_m)
\geq t_1 \}$ and $\cV_2 := \{ \bu_r: S_2(\bu_r) \geq t_2 \}$
are created. Let $q_i$ be the probability mass function for
$\cV_i$. A word bank for $\cV$ can then be created by repeating the following steps.

\begin{enumerate}
\item Select $\bv_m$ and $\bu_r$ independently from their
respective word banks.

\item Select $d$ randomly from $\{ d_1, \ldots, d_2 \}$. Generate
$z_1, \ldots, z_d$ sequentially from the underlying Markov chain, initialized
at $z_0 = v_m$.

\item Store $\bw = \bv_m \bz_d \bu_r$ into the word bank.
\end{enumerate}

Let $q$ be the probability mass function of the stored words. Then
\begin{equation} \label{qw} q(\bw) = (d_2-d_1+1)^{-1} q_1(\bv_m)
\sigma(v_m \bz_d) q_2(\bu_r)
\end{equation}
and hence $\beta(\bw)=q(\bw)/\sigma(\bw)=(d_2-d_1+1)^{-1} \beta_1(\bv_m)
\beta_2(\bu_r)/\sigma(z_d u_1)$.

\smallskip \noindent
{\it 5.1 Example 3}

\smallskip The transcription factors SFF (with PSWM $W_1$)
and MCM1 (with PSWM $W_2$)
are regulators of the cell cycle in yeast, and are known to co-operate at close distance
in the promoter regions of the genes they regulate,
see Spellman et al. (1998). Their PSWMs can be obtained from the database SCPD. Define $\cV$ by
(\ref{v31}) with $t_1=48$, $t_2=110$, $d_1=0$ and $d_2=100$. We would like to estimate the
probability that the motif $\cV$ appears at least once
within a promoter sequence of length $n=700$. The estimated probability using Algorithm A is
$3.4 \times 10^{-3}$ with a standard error of $3 \times 10^{-4}$. The corresponding standard
error for 1000 direct Monte Carlo runs would have been about $2 \times 10^{-3}$, which is
large relative to the underlying probability.

\smallskip
\noindent {\it 5.2 Structured Motifs}

\smallskip These co-occurring motifs considered in Robin et al. (2002)
consist essentially of fixed word patterns $\bx_m$ and
$\by_r$ separated by a gap of length $d$,
with an allowance for the mutation of up to one base in
$\bx_m \by_r$. The motif can be expressed as
\begin{equation}
\label{wzy}
\cV = \{ \bv_m \bz \bu_r: d_1 \leq \ell(\bz) \leq d_2, |\{ i: v_i \neq x_i \}|
+ | \{ i: u_i \neq y_i \}| \leq 1 \}.
\end{equation}
We create a word for the word bank of $\cV$ in the following manner.

\begin{enumerate}
\item Select $k$ randomly from $\{ 0, \ldots, m+r \}$. If $k=0$, then there is
no mutation and we let $\bv_m \bu_r = \bx_m \by_r$. Otherwise, change the $k$th base of
$\bx_m \by_r$ equally likely into one of the three other bases and
denote the mutated sequence as $\bv_m \bu_r$.

\item Select $d$ randomly from $\{ d_1, \ldots, d_2 \}$ and generate the bases of
$\bz=z_1 \cdots z_d$ sequentially from
the underlying Markov chain, initialized at $z_0=v_m$.
\end{enumerate}

We perform a simulation study on eight structural motifs
selected for their high frequency of occurrences in part of the
{\it Bacillus subtilis} DNA dataset. We consider
$(d_1,d_2) = (16,18)$ and $(5,50)$, with length of DNA
sequence $n=100$, and a Markov chain with transition matrix
$$\begin{matrix} a \cr c \cr g \cr t \end{matrix} \left( \begin{array}{cccc}
0.35 & 0.16 & 0.18 & 0.31 \cr
0.33 & 0.20 & 0.15 & 0.32 \cr 0.32 &
0.22 & 0.19 & 0.27 \cr 0.25 & 0.20 &
0.19 & 0.35 \end{array} \right).
$$
In Table 3, we compare importance sampling estimates of $P \{ N \geq 1 \}$ using
Algorithm A with analytical p-value estimates
from Robin et al. (2002) and direct Monte Carlo p-value estimates. The analytical p-value
estimates are computed numerically
via recursive methods with computation time that grows exponentially
with $d_2-d_1$, and are displayed only for the case $(d_1,d_2)=(16,18)$.

We illustrate here how
the importance sampling algorithms can be modified to handle more complex situations, for
example, to obtain a combined p-value for all eight motifs. Consider more generally
$p=P \{ \max_{1 \leq j \leq J} (N^{(j)}-c_j) \geq
0 \}$, where $N^{(j)}$ is the
total word count from the motif $\cV^{(j)}$ and $c_j$ is a positive integer. Let $L^{(j)}$
be the likelihood ratio when applying either Algorithm A or B with insertion of words from
$\cV^{(j)}$. For the
$k$th simulation run, we execute the following steps.

\begin{enumerate}
\item Select $j_k$ randomly from $\{ 1,\ldots, J \}$.

\item Generate $\bs^{(k)}_0$ using either Algorithm
A or B, with insertion of words from $\cV^{(j)}$.
\end{enumerate}

\smallskip \noindent Then
\begin{equation}
\label{combined}
\wht p_I = K^{-1} \sum_{k=1}^K [L^{(j_k)}(\bs^{(k)}_0)]^{-1}
\Big( \frac{J}{|\{ j: N^{(j)}(\bs^{(k)}_0) \geq c_j \}|} \Big) {\bf 1}_{\{
N^{(j_k)}(\bs^{(k)}_0) \geq c_{j_k} \}}
\end{equation}
is unbiased for $p$, see Appendix C. The key feature in (\ref{combined})
is the correction term $|\{ j: N^{(j)}(\bs^{(k)}_0) \geq c_j \}|$.
Without this term, $\wht p_I$ is an unbiased estimator for the
Bonferroni upper bound $\sum_{j=1}^J P \{ N^{(j)} \geq c_j \}$.
The correction term adjusts the estimator downwards when more than one thresholds
$c_j$ are exceeded.

We see from Table 3 that the variance reduction is substantial when importance sampling
is used. In fact, the direct Monte Carlo estimate is often
unreliable. Such savings in computation time is valuable both to the
end user and also to the researcher trying to test the reliability
of his or her analytical estimates on small p-values.
We observe for example that
the numerical estimates for $(d_1,d_2)=(16,18)$
given in Robin et al. (2002) are quite accurate but tends to
underestimate the true underlying probability.

\newpage

\def\theequation{6.\arabic{equation}}
\setcounter{equation}{0}
\begin{center}
{\bf 6. DISCUSSION}
\end{center}
\smallskip \noindent
The examples given here are not meant to be exhaustive but they do indicate how we can proceed
in situations not covered here. For example, if we would like the order of the two words in a
CRM to be arbitrary, we can include an additional permutation step in the construction of the
word bank. In Section 5.2, we also showed how to simulate p-values of the maximum count over a
set of word patterns. As we gain biological understanding, the models that we formulate for DNA
and protein functional sites become more complex. Over the years, they have evolved from
deterministic words to consensus sequences to PSWMs and then to motif modules. As probabilistic
models for promoter architecture gets more complex and context specific, importance sampling
methods are likely to be more widely adopted in the computation of p-values.

\bigskip
\begin{center}
{\bf ACKNOWLEDGMENTS}
\end{center}

\smallskip This research was partially supported by National University of Singapore
grants C-389-000-010-101 and R-155-062-112.

\bigskip
\begin{center}
{\bf DISCLOSURE STATEMENT}
\end{center}

\smallskip No competing financial interests exist.

\newpage
\def\theequation{A.\arabic{equation}}
\setcounter{equation}{0}
\begin{center}
{\bf APPENDIX A}
\end{center}

\smallskip
We first show how words $\bv_m$ can be generated with probability mass function
$$q(\bv_m) = \pi(v_1) \sigma(\bv_m) \sigma(\bv_m^c)/\eta,
$$
with $\eta = \sum_{\bv_m} \pi(v_1) \sigma(\bv_m) \sigma(\bv_m^c)$ a computable
normalizing constant.
Apply the backward recursive relations
\begin{equation} \label{etai}
\eta_i(x) = \sum_{y \in \cX} \sigma(xy) \sigma(y^c x^c) \eta_{i+1}(y)
\mbox{ for all } x \in \cX \mbox{ and } i=1,\ldots,m-1,
\end{equation}
initialized with $\eta_m(x) = 1$ for all $x$.
Then $\eta = \sum_{x \in \cX} \pi(x) \eta_1(x)$. Let $Q$ be the desired probability
measure for generating $\bv_m$ with probability mass function $q$. Then the Markovian property
\begin{eqnarray} \label{Qv1}
Q \{ v_1=x \} & = & \pi(x) \eta_1(x)/\eta, \cr
Q \{ v_{i+1}=y|v_i=x \} &=& \sigma(xy) \sigma(y^c x^c) \eta_{i+1}(y)/\eta_i(x)
\mbox{ for } i=1,\ldots,m-1,
\end{eqnarray}
allows us to generate $v_i$ sequentially via transition matrices.

To generate words $\bv_m$ with probability mass function $q(\bv_m)=\pi(v_1) \sigma(\bv_m
\bv_m^c)/\eta$, let $\eta_m(x)=\sigma(xx^c)$ instead of $\eta_m(x)=1$ and proceed
with (\ref{etai}) and (\ref{Qv1}).

\bigskip
\begin{center}
{\bf APPENDIX B}
\end{center}

\smallskip
Let $S$ be the score with respect to a
given PSWM $W$ and let $\theta > 0$. We provide here
a quick recursive algorithm for generating $\bv_m$ from
the probability mass function
\begin{equation}
\label{qve}
q_\theta(\bv_m) = e^{\theta S(\bv_m)} \pi(v_1) \sigma(\bv_m)/\Lambda(\theta),
\end{equation}
with $\Lambda(\theta)=\sum_{\bv_m} e^{\theta S(\bv_m)} \pi(v_1) \sigma(\bv_m)$ a computable normalizing
constant. Since $\log \Lambda(\theta)$ is convex, the
solution of $\frac{d}{d \theta} [\log \Lambda(\theta)]=t$ can be found using a bijection search.
We take note of the backward recursive
relations
\begin{eqnarray}
\label{Llt}
\Lambda_m(\theta,x) & = & e^{\theta w_m(x)}, \cr
\Lambda_i
(\theta,x) & = & e^{\theta w_i(x)} \sum_{y \in \cX} \sigma(xy) \Lambda_{i+1}
(\theta,y) \mbox{ for all } x \in \cX \mbox{ and } i=1,\ldots,m-1,
\end{eqnarray}
from which we can compute $\Lambda(\theta)=\sum_{x \in \cX} \pi(x) \Lambda_1(\theta,x)$.
Let $Q$ denote the desired probability measure for generating $\bv_m
=v_1 \cdots v_m$ from $q_\theta$. By (\ref{qve}) and (\ref{Llt}), we
can simply generate the letters $v_i$ sequentially, using transition
matrices defined by the Markovian relations
\begin{eqnarray}
\label{Pv1}
Q \{ v_1=x \} & = & \pi(x) \Lambda_1(\theta,x)/\Lambda(\theta), \cr
Q \{ v_{i+1}=y|v_i=x \} &=& e^{\theta w_i(x)} \sigma(xy) \Lambda_{i+1}(\theta,y)/\Lambda_i
(\theta,x) \mbox{ for } i=1,\ldots,m-1.
\end{eqnarray}

\bigskip
\begin{center}
{\bf APPENDIX C}
\end{center}

\smallskip We shall show here that $\wht p_I$ in (\ref{combined}) is unbiased for
$p=P \{ \max_{1 \leq j \leq J} (N^{(j)}-c_j) \geq 0 \}$. Let $A_j = \{ \bs_0: N^{(j)}(\bs_0)
\geq c_j \}$ and let $Q_j$ be a probability measure such that
$L^{(j)}(\bs_0)=Q_j(\bs_0)/P(\bs_0) > 0$ for any $\bs_0 \in A_j$. Let $A=\cup_{j=1}^J
A_j$. Then with the convention $0/0=0$,
$$J^{-1} \sum_{j=1}^J E_{Q_j} \Big\{ [L^{(j)}(\bs_0)]^{-1} \Big( \frac{J}{| \{ \ell:
\bs_0 \in A_\ell \}|} \Big) {\bf 1}_{\{ \bs_0 \in A_j \}} \Big\} = E \Big(
\frac{\sum_{j=1}^J {\bf 1}_{\{ \bs_0 \in A_j \}}}{| \{ \ell: \bs_0 \in A_\ell \}|}
\Big) = P \{ \bs_0 \in A \},
$$
and hence $\wht p_I$ is indeed unbiased.

\newpage
\bigskip
\begin{center}
{\bf APPENDIX D: ASYMPTOTIC OPTIMALITY}
\end{center}
To estimate $p:=P \{ N(\bs) \geq c \}$ using direct Monte Carlo, simply generate
$K$ independent copies of $\bs$, denoted by $\bs^{(1)},\ldots,\bs^{(K)}$, under the original
probability
measure $P$, and let
$$\wht p_{\rm D}= K^{-1} \sum_{k=1}^K {\bf 1}_{\{ N(\bs^{(k)}) \geq c \}}.
$$
To simulate $p$ using importance sampling, we need to first
select a probability measure $Q \neq P$
for generating $\bs^{(1)},\ldots,\bs^{(K)}$. The estimate of $p$ is then
$$\wht p_{\rm I}:= K^{-1} \sum_{k=1}^K L^{-1}(\bs^{(k)}) {\bf 1}_{\{ N(\bs^{(k)}) \geq c \}},
\mbox{ where } L(\bs) =Q(\bs)/P(\bs).
$$
We require $Q(\bs)>0$
whenever $N(\bs) \geq c$, so as to ensure that $\wht p_{\rm I}$ is unbiased for $p$.

The relative error (RE) of a Monte Carlo estimator $\wht p=\wht p_{\rm D}$ or
$\wht p_{\rm I}$, is given
by $\sqrt{\var(\wht p)}/p$. We say that
$\wht p$ is asymptotically optimal if for any $\epsilon > 0$, we can satisfy
RE $\leq \epsilon$ with
$\log K=o(| \log p|)$ as $p \rightarrow 0$, see Sadowsky and Bucklew (1990) and
Dupuis and Wang (2005). Since RE$(\wht p_{\rm D})=\sqrt{(1-p)/(Kp)}$, direct
Monte Carlo is not asymptotically optimal. The question we would like to answer here is:
Under what conditions
are Algorithms A and B asymptotically optimal?

The examples described in Sections 3--5 involve
word families that can be characterized as $\cV_m$. We may also include an additional
subscript $m$ to a previously defined quantity to highlight its dependence on $m$, for example
$p_m,q_m,\beta_m$ and $n_m$. We say that $x_m$ and $y_m$ have similar logarithmic value relative
to $m$, and write $x_m \simeq y_m$, if $|\log x_m - \log y_m|=o(m)$ as $m \rightarrow \infty$. It
is not hard to see that if $x_m \simeq y_m$ and $y_m \simeq z_m$, then
$x_m \simeq z_m$. In Algorithm A, it is assumed implicitly that
$n_m \geq \ell_{\max} (=\ell_{\max,m})
:=\max_{\bv \in \cV_m} \ell(\bv)$ and we shall also assume $n_m \geq c
\ell_{\max}$ when using Algorithm B. To fix the situation, let $\rho_i=c/n_m$ for all $i$
in Algorithm B. Let $\beta_{\min}(=\beta_{\min,m})=
\min_{\bv \in \cV_m} \beta_m(\bv)$, $\beta_{\max}(=\beta_{\max,m})
=\max_{\bv \in \cV_m} \beta_m(\bv)$,
$\sigma_{\min} = \min_{x,y \in \cX} \sigma(xy) (>0)$, $\sigma_{\max} =
\max_{x,y \in \cX} \sigma(xy) (<1)$ and $\pi_{\min} =\min_{x \in \cX} \pi(x)
(\geq \sigma_{\min})$. Let
$\lfloor \cdot \rfloor$ denote the greatest integer function, $P_x$ denote probability
conditioned on $s_1=x$ or $v_1=x$ and $P_\pi$ denote probability conditioned on $s_1$ or $v_1$
following the stationary distribution.

In the following lemma, we provide conditions for asymptotic optimality and check them
in Appendices D.1--D.3
for the word families discussed in Sections 3--5.

\medskip {\bf Lemma 1}. {\it If $\log n_m \simeq 1$ and
\begin{eqnarray} \label{pm}
p_m & \leq & \alpha^m \mbox{ for some } 0 < \alpha < 1, \\
\label{condl}
\ell_{\max} & \simeq & 1, \\
\label{betamin} \beta_{\min} & \simeq & \Big( \sum_{\bv \in \cV_m}
\sigma(\bv) \Big)^{-1},
\end{eqnarray}
then both Algorithms A and B are asymptotically optimal.}

\medskip {\it Proof}. Let $r_m = \sum_{x \in \cX}
P_x \{ \bs_\ell \in \cV_m$ for some $\ell \geq 1 \}$. Since $\sum_{\bv \in \cV_m} \sigma(\bv)
\geq r_m \geq \ell_{\max}^{-1} \sum_{\bv \in \cV_m} \sigma(\bv)$, by (\ref{condl}) and
(\ref{betamin}),
\begin{equation} \label{rm}
r_m \simeq \sum_{\bv \in \cV_m} \sigma(\bv) \simeq \beta_{\min}^{-1}.
\end{equation}
By (6.1),
$|\log p_m| \geq m | \log \alpha |$ for all large $m$ and hence
it suffices for us to show $K_m \simeq 1$.

If $n_m \simeq 1$, then by (\ref{rm}) and the inequalities
${n_m \choose c} r_m^c \geq p_m \geq (\sigma_{\min} r_m)^c$,
\begin{equation} \label{nmrmc}
(n_m \beta_{\min}^{-1})^c \simeq (n_m r_m)^c \simeq p_m.
\end{equation}
Consider next the case $n_m/\ell_{\max} \rightarrow \infty$. Since $\log n_m \simeq 1$, there
exists integers $\xi_m$ such that $\xi_m \simeq 1$, $\xi_m=o(n_m)$ and $\log n_m = o(\xi_m)$.
Let $\kappa_m = \lfloor n_m/(\ell_{\max}+\xi_m) \rfloor$ and
$g_m =P_\pi \{ s_\ell \in \cV_m$ for some $\ell \geq 1 \}$. By (\ref{pm}),
$\alpha^m \geq p_m \geq (g_m \sigma_{\min})^c$ and
hence $g_m \rightarrow 0$. Since the underlying Markov chain is uniformly ergodic,
\begin{equation} \label{uerg}
\sup_{x,y \in \cX} | P_x \{ s_{k+1}=y \}-\pi(y) | \leq \eta^k \mbox{ for some } 0 < \eta < 1.
\end{equation}
By considering the sub-cases
of at least $c$ words $\bv \in \cV_m$ starting at positions $1, (\ell_{\max}+\xi_m)
+1, \ldots, (\kappa_m-1)(\ell_{\max}+\xi_m)+1$, it follows from (\ref{uerg}) that
$$p_m \geq 1- \sum_{j=0}^{c-1} {\kappa_m \choose j}
g_m^j (1-g_m)^{\kappa_m-j} -(\kappa_m-1) \eta^{\xi_m} =
1-(1+o(1)) \sum_{j=0}^{c-1} \frac{(\kappa_m g_m)^j}{j !}  e^{-\kappa_m g_m}-o(1).
$$
By (\ref{pm}), $\kappa_m g_m \rightarrow 0$
and this implies $\kappa_m r_m \rightarrow 0$. Since $(\ell_{\max}+\xi_m) \simeq 1$, it follows
that $\kappa_m \simeq n_m$ and hence by the inequalities
$${n_m \choose c} r_m^c \geq p_m \geq {\kappa_m \choose c}
(\sigma_{\min} r_m)^c (1-r_m)^{\kappa_m-c},
$$
(\ref{nmrmc}) again holds. By using a subsequence argument if necessary, it follows that
(\ref{nmrmc}) holds as long as $\log n_m \simeq 1$.

For Algorithm A, by (\ref{Lsk}) and (\ref{pI}),
$$\mbox{RE}(\wht p_{\rm I}) \leq p_m^{-1} K_m^{-1/2} \sup_{\bs} L^{-1}(\bs) {\bf 1}_{\{ N(\bs)
\geq 1 \}} \leq p_m^{-1} K_m^{-1/2} n_m \sigma_{\max} \beta_{\min}^{-1}
$$
and the
desired relation
$K_m \simeq 1$ follows from (\ref{nmrmc}) with $c=1$.

For Algorithm B, it follows from (\ref{Lis}) that if $N(\bs) \geq c$, then
$L(\bs) \geq (1-c/n_m)^{n_m}$ $\times [c \beta_{\min}/(n_m \sigma_{\max})]^c$
and hence by (\ref{pI}),
$$\mbox{RE}(\wht p_{\rm I}) \leq p_m^{-1} K_m^{-1/2} \sup_{\bs} L^{-1}(\bs) {\bf 1}_{\{
N(\bs) \geq c \}} \leq (1+o(1)) p_m^{-1} K_m^{-1/2}[e n_m \sigma_{\max}/(c \beta_{\min})]^c,
$$
and again $K_m \simeq 1$ follows from (\ref{nmrmc}). $\Box$

\smallskip
\noindent {\it D.1 Inverted repeats}

\smallskip
Consider the word family (\ref{invert}) with $d_2 \simeq 1$. Then (\ref{condl}) holds.
Since $p_m \leq (d_2-d_1) n_m \sigma_{\max}^{2m-1}$, (\ref{pm})
holds when $n_m=O(\gamma^m)$ for some $\gamma < \sigma_{\max}^{-2}$. It remains to check
(\ref{betamin}). Since
$\sum_{\bv \in \cV_m} q_m(\bv)=\sum_{\bv \in \cV_m} \beta_m(\bv) \sigma(\bv)=1$,
\begin{equation} \label{betabound}
\beta_{\min} \leq \Big( \sum_{\bv \in \cV_m} \sigma(\bv) \Big)^{-1} \leq \beta_{\max}.
\end{equation}
Let $\bu_m$ be generated with
probability proportional to $\pi(u_1) \sigma(\bu_m) \sigma(\bu_m^c)$ when creating the
word bank $\cV_m$. Then there exists a constant $C>0$ such that
$$\beta_m(\bu_m \bz \bu_m^c) = C \pi(u_1) \sigma(\bu_m \bz) \sigma(\bu_m^c)/
\sigma(\bu_m \bz \bu_m^c) = C \pi(u_1)/\sigma(z_d u_1^c).
$$
Hence $\beta_{\min} \simeq \beta_{\max}$ and (\ref{betamin})
follows form (\ref{betabound}).

\smallskip
\noindent {\it D.2 Word patterns derived from PSWMs}

\smallskip For the word family (\ref{VPSWM}), condition (\ref{condl}) is always
satisfied. Let the entries of
the PSWM be non-negative integers and assume that the column totals are fixed at some
$C>0$. It follows from large deviations theory, see for example Dembo and Zeitouni (1998),
that if $t(=t_m) \geq E_\pi S(\bv)+\zeta m$
for some $\zeta > 0$, then
\begin{equation} \label{ld}
P_\pi \{ S(\bv) \geq t \} = O(\lambda^m) \mbox{ for some } 0 < \lambda < 1.
\end{equation}
 Since $p_m \leq n_m P_\pi \{ S(\bv) \geq t \}$,
(\ref{pm}) holds if $n_m = O(\gamma^m)$ for some $\gamma < \lambda^{-1}$.

To simplify the analysis in checking (\ref{betamin}), select the
tilting parameter $\theta(=\theta_m)$ to be
the root of $E_{q_\theta}[S(\bv)]
=t+\delta_m$ for some positive $\delta_m=o(m)$ satisfying $m^{-1/2}
\delta_m \rightarrow \infty$ as $m \rightarrow \infty$, instead of the root of $E_{q_\theta}
[S(\bv)]
=t$, as suggested in the statement containing (\ref{qstar}). The implicit assumption is that
$\sum_{i=1}^m \{ \max_{v \in \cX} w_i(v) \} > t+\delta_m$ for all $m$. Since the entries of
the transition matrices derived in Appendix B are uniformly bounded away from zero, it follows
from a coupling argument that $\cov_{q_\theta}(w_i(v_i),w_j(v_j)) = O(\tau^{|i-j|})$ for some
$0 < \tau < 1$ and hence $\var_{q_\theta} (S(\bv)) =O(m)$.
By (\ref{tilted}) and Chebyshev's inequality,
\begin{equation} \label{sumsig}
e^{\theta(t+2 \delta_m)} \sum_{\bv \in \cV_m} \sigma(\bv) \Big/ \Lambda(\theta) \geq
\sum_{\bv: |S(\bv) - t- \delta_m| \leq \delta_m} q_\theta(\bv) \geq 1- \delta_m^{-2}
\var_\theta (S(\bv)) > 0
\end{equation}
for all large $m$. Since $\xi>1$ in (\ref{tq}), $\beta_{\min} =\min_{\bv \in \cV_m}
q_m(\bv)/\sigma(\bv) > e^{\theta t} \pi_{\min}/\Lambda(\theta)$ and (\ref{betamin})
follows from (\ref{betabound}) and (\ref{sumsig}).

\smallskip
\noindent {\it D.3 Co-occurrences of motifs}

\smallskip Consider the word family (\ref{v31}) with $(r/m)$ bounded away from zero and
infinity and $d_2 \simeq 1$. We check that (\ref{condl}) holds. If $t_1 \geq ES_1(\bv) + \zeta m$
for some $\zeta>0$, then (\ref{ld}) holds
with $S$ replaced by $S_1$, $t$ replaced by $t_1$ and hence (\ref{pm}) holds if $n_m=
O(\gamma^m)$ for some $\gamma < \lambda^{-1}$.

Let $\theta_j$ be the root of $E_{\theta_j}[ S_j(\bv) ]= t_j+\delta_m$
for some positive $\delta_m=o(m)$ with $m^{1/2} \delta_m \rightarrow \infty$, $j=1$ and 2,
assuming that $\sum_{i=1}^{m_j} \{ \max_{v \in \cX}
w_i^{(j)}(v) \} > t_j+\delta_m$, where
$m_1=m$ and $m_2=r$. Let $\cV_m^{(1)} = \{ \bv_m: S_1(\bv_m)
\geq t_1 \}$, $\cV_r^{(2)} = \{ \bu_r: S_2(\bu_r) \geq t_2 \}$ and let $\Lambda^{(1)}(\theta_1)$,
$\Lambda^{(2)}(\theta_2)$ be their respective normalizing constants, see (\ref{tilted}).
By the arguments in
(\ref{sumsig}),
$$\sum_{\bv \in \cV_m} \sigma(\bv) \geq \sigma_{\min} \Big( \sum_{\bv \in \cV_m^{(1)}}
\sigma(\bv) \Big) \Big( \sum_{\bu \in \cV_r^{(2)}} \sigma(\bu) \Big) =e^{-\theta_1 t_1-\theta_2
t_2+o(m)} \Lambda^{(1)}(\theta_1) \Lambda^{(2)}(\theta_2).
$$
By (\ref{qw}), $\beta_{\min} \geq e^{\theta_1 t_1+\theta_2 t_2} d_2^{-1} \pi^2_{\min}/
\{ \Lambda^{(1)}(\theta_1) \Lambda^{(2)}(\theta_2) \}$ and hence (\ref{betamin}) follows from
(\ref{betabound}).

\newpage
\renewcommand{\ref}{\par\noindent\hangindent 12pt}
\begin{center}
{\bf REFERENCES}
\end{center}
\medskip
\ref Chan, H.P. and Zhang, N.R. 2007. Scan statistics with weighted
observations. {\it J. Am. Statist. Ass.}, 102, 595--602.

\ref Chiang, D.Y., Moses, A.M., Kellis, M., Lander, E. and Eisen, M.
2003. Phylogenetically and spatially conserved word pairs
associated with gene-expression changes in yeasts.  \emph{Genome
Biol.}, 4, R43.

\ref Cottrell, M., Fort, J.C. and Malgouyres, G. 1983.
Large deviations and rare events in the study of stochastic algorithms.
{\it IEEE Trans. Automat. Contr.}, 28, 907--920.

\ref Dembo, A. and Zeitouni, O. 1998. {\it Large deviations techniques and applications}.
Springer, New York.

\ref Do, K.A. and Hall, P. 1992. Distribution estimating using concomitant
of order statistics, with applications to Monte Carlo simulation
for the bootstrap, {\it J.R. Statist. Soc.} B, 54, 595--607.

\ref Dupuis, P. and Wang, H. 2005. Dynamic importance sampling for uniformly
recurrent Markov chains. {\it Ann. Appl. Probab}, 15, 1--38.

\ref Fuh, C.D. and Hu, I. 2004. Efficient importance sampling for events
of moderate deviations with applications. {\it Biometrika}, 91, 471--490.

\ref Gusfield, D. 1997. {\it Algorithms on Strings, Trees and Sequences:
Computer Science and Computational Biology}. Cambridge University Press, London.

\ref Huang, H., Kao, M., Zhou, X., Liu, J., and Wong, W. 2004.
Determination of local statistical
significance of patterns in Markov sequences with applications to
promoter element identification. {\it J. Comput. Biol.}, 11, 1--14.

\ref Johns, M.V. 1988. Importance sampling for bootstrap confidence
intervals, {\it J. Am. Statist. Ass.}, 83, 709--714.

\ref Lai, T.L. and Shan, J.Z. 1999.
Efficient recursive algorithms for detection of abrupt
changes insignals and control systems. {\it IEEE Trans. Automat. Contr.}, 44, 952--966.

\ref Leung M.Y., Choi K.P., Xia A. and Chen, L.H.Y. 2005.
Nonrandom clusters of palindromes in herpesvirus genomes. {\it J. Comput. Biol.}, 12, 331--354.

\ref Leung M.Y., Schachtel G.A. and Yu H.S.
1994. Scan statistics and DNA sequence analysis:
The search for an origin of replication in a virus. {\it Nonlinear World}.

\ref Masse, M.J.O., Karlin, S., Schachtel, G.A. and Mocarski, E.S. 1992.
Human cytomegalovirus origin of DNA replication (oriLyt) resides within a highly complex
repetitive region. {\it Proc. Natn Acad. Sci.}, 89, 5246--5250.

\ref Mitrophanov, A.Y. and Borodovsky, M. 2006.
Statistical significance in biological sequence analysis.
{\it Briefings Bioinformatics}, 7, 2--24.

\ref Pape, U., Rahmann, S., Sun, F. and Vingron, M. 2008. Compound Poisson
approximation of the number of occurrences of a position frequency matrix (PFM) on both
strands. {\it J. Comput. Biol.}, 15, 547--564.

\ref Prum, B., Rodolphe, F. and de Turckheim, E. 1995. Finding words
with unexpected frequencies in deoxyribonucleic acid sequences. {\it J.R.
Statist. Soc.} B, 57, 205--220.

\ref R\'{e}gnier, M. 2000. A unified approach to word occurrence
probabilities. {\it Dis. Appl. Math.}, 104, 259--280.

\ref Reinert, G., Schbath, S. and Waterman, M. 2000. Probabilistic
and statistical properties of words: An overview. {\it J. Comput. Biol.}, 7,
1--46.

\ref Robin, S., Daudin, J., Richard, H., Sagot, M. and Schbath, S. 2002. Occurrence
probability of structured motifs in random sequences. {\it J. Comput. Biology},
9, 761--773.

\ref Sadowsky, J.S. and Bucklew, J.A. 1990. On large deviations theory and
asymptotically efficient Monte Carlo estimation. {\it IEEE Trans. Info. Theory},
36, 579--588.

\ref Siegmund, D. 1976. Importance sampling in the Monte Carlo study of sequential
test. {\it Ann. Statist.}, 4, 673--684.

\ref Spellman P.T., Sherlock, G., Zhang, M.Q., Iyer, V.R., Anders, K., Eisen, M.B., Brown,
P.O., Botstein, D., Futcher, B.
1998.  Comprehensive Identification of Cell Cycle-regulated Genes of the Yeast Saccharomyces cerevisiae by Microarray Hybridization.  {\it Molecular Biology of the Cell}, 9, 3273--3297.

\ref Zhang, N.R., Wildermuth, M.C. and Speed, T.P. 2008.
Transcription factor binding site prediction with multivariate gene
expression data. {\it Ann. Appl. Statist.}, 2, 332--365.

\ref Zhou, Q. and Wong, W. 2004. CisModule: De novo
discovery of cis-regulatory modules by hierarchical mixture
modeling. {\it Proc. Natn Acad. Sci.}, 101, 12114--112119.

\ref Zhu, J. and Zhang, M.Q. 1999. SCPD: a promoter database of the yeast Saccharomyces
cerevisiae. \emph{Bioinformatics}, 15, 607--611.

\newpage
\begin{table}
\begin{center}
\begin{tabular}{r|r|r|r}
$t$ & 9 & 10 & 11 \cr \hline
Analytical & 7.1$\times 10^{-2}$ & 7.1$\times 10^{-3}$ & 4.2$\times 10^{-4}$ \cr \hline
$W_{\rm rep}$: Direct MC & $(3.6\pm.6) \times 10^{-2}$ & $(5\pm2) \times 10^{-3}$
& 0 \cr
Algorithm A & $(3.0\pm.1) \times 10^{-2}$ & $(4.0\pm.2)\times 10^{-3}$ &
$(2.7\pm.1) \times 10^{-4}$ \cr \hline
$W_{\rm norep}$: Direct MC & $(6.7\pm.8) \times 10^{-2}$ & $(9\pm3) \times 10^{-3}$
& $(1 \pm 1) \times 10^{-3}$ \cr
Algorithm A & $(7.5\pm.2) \times 10^{-2}$ & $(6.9\pm.2) \times 10^{-3}$ &
$(4.1\pm.1) \times 10^{-4}$ \cr
\end{tabular}
\end{center}
\caption{ Comparisons of analytical, direct Monte Carlo and importance sampling
approximations for $P \{ N \geq 1 \}$ with $n=200$ in Example 1.
Each Monte Carlo entry is obtained using 1000 simulation runs and
are expressed in the form $\wht p \pm$ standard error.}
\end{table}

\begin{table}
\begin{center}
\begin{tabular}{r|r|r}
$c$ & {\rm Direct MC} & {\rm Algorithm B} \cr \hline
1 & (9.6$\pm.9) \times 10^{-2}$ & (9.1$\pm.3) \times 10^{-2}$ \cr
2 & (3$\pm2) \times 10^{-3}$ & (4.2$\pm.2) \times 10^{-3}$ \cr
3 & 0 & $(1.3 \pm .1) \times 10^{-4}$ \cr
4 & 0 & $(2.6 \pm .3) \times 10^{-6}$ \cr
\end{tabular}
\end{center}
\caption{ $\widehat p \pm$standard error for Example 2
with 1000 copies of $\bs_0$ generated for both direct Monte Carlo and importance sampling using
Algorithm B.}
\end{table}

\begin{table}
\begin{center}
\begin{tabular}{rr|rr|r|r|r}
$d_1$ & $d_2$ & $\bx$ & $\by$ & Direct MC &
Algorithm A & Analytic \cr
\hline
16 & 18 & $gttgaca$ & $atataat$ & $(2 \pm 1) \times 10^{-4}$ &
$(1.038 \pm 0.006) \times 10^{-4}$ & $1.01 \times 10^{-4}$ \cr
& & $gttgaca$ & $tataata$ & 0 & $(9.00 \pm 0.05) \times 10^{-5}$
& $8.82 \times 10^{-5}$ \cr
& & $tgttgac$ & $tataata$ & $(20 \pm 10) \times 10^{-5}$ &
$(9.39 \pm 0.05) \times 10^{-5}$ & $9.20 \times 10^{-5}$ \cr
& & $ttgaca$ & $ttataat$ & $(9 \pm 3) \times 10^{-4}$ &
$(6.65 \pm 0.03) \times 10^{-4}$ & $6.55 \times 10^{-4}$ \cr
& & $ttgacaa$ & $tacaat$ & $(4 \pm 2) \times 10^{-4}$ &
$(4.64 \pm 0.02) \times 10^{-4}$ & $4.57 \times 10^{-4}$ \cr
& & $ttgacaa$ & $tataata$ & $(2 \pm 1) \times 10^{-4}$ &
$(1.798 \pm 0.009) \times 10^{-4}$ & $1.78 \times 10^{-4}$ \cr
& & $ttgacag$ & $tataat$ & $(5 \pm 2) \times 10^{-4}$ &
$(3.62 \pm 0.02) \times 10^{-4}$ & $3.59 \times 10^{-4}$ \cr
& & $ttgacg$ & $tataat$ & $(10 \times 3) \times 10^{-4}$ &
$(9.90 \pm 0.06) \times 10^{-4}$ & $9.76 \times 10^{-4}$ \cr
& & \multicolumn{2}{c|}{combined p-value} & $(2.0 \pm 0.4)
\times 10^{-3}$ & $(2.96 \pm 0.03) \times 10^{-3}$ \cr
\hline
5 & 50 & $gttgaca$ & $atataat$ & $(1 \pm 0.3) \times 10^{-3}$ &
$(1.265 \pm 0.008) \times 10^{-3}$ \cr
& & $gttgaca$ & $tataata$ & $(0.4 \pm 0.2) \times 10^{-3}$ &
$(1.103 \pm 0.007) \times 10^{-3}$ \cr
& & $tgttgac$ & $tataata$ & $(1.8 \pm 0.4) \times 10^{-3}$ &
$(1.150 \pm 0.007) \times 10^{-3}$ \cr
& & $ttgaca$ & $ttataat$ & $(7.4 \pm 0.9) \times 10^{-3}$ &
$(7.88 \pm 0.05) \times 10^{-3}$ \cr
& & $ttgacaa$ & $tacaat$ & $(5.0 \pm 0.7) \times 10^{-3}$ &
$(5.50 \pm 0.04) \times 10^{-3}$ \cr
& & $ttgacaa$ & $tataata$ & $(1.5 \pm 0.4) \times 10^{-3}$ &
$(2.21 \pm 0.01) \times 10^{-3}$ \cr
& & $ttgacag$ & $tataat$ & $(3.1 \pm 0.6) \times 10^{-3}$ &
$(4.23 \pm 0.03) \times 10^{-3}$ \cr
& & $ttgacg$ & $tataat$ & $(0.9 \pm 0.1) \times 10^{-2}$ &
$(1.126 \pm 0.008) \times 10^{-2}$ \cr
& & \multicolumn{2}{c|}{combined p-value} & $(2.7 \pm 0.2) \times
10^{-2}$ &
$(3.30 \pm 0.04) \times 10^{-2}$ \cr
\end{tabular}
 \caption{ Comparison of direct Monte Carlo,
importance sampling  and analytical estimates of $P \{ N \geq 1 \}$ for structured motifs.
For both direct Monte Carlo and importance sampling, 10,000 simulation runs are
executed for each entry and the results are displayed in the form $\wht p \pm$standard
error.}
\end{center}
\end{table}

\end{document}